# Asymmetry in Political Geography and Compactness in Districting: a Computational Analysis of Bias

Constantine (Dinos) Gonatas

*CPG Advisors, Concord MA USA*

cpgonatas@cpg-advisors.net

260 Old Marlboro Rd, Concord MA USA

**Abstract**

We investigate the distribution of partisanship in a cross-section of ten diverse States to elucidate how votes translate into seats won and other metrics. Markov chain simulations taking into account partisanship distribution agree surprisingly well with a simple model covering only equal voting population-weighted distributions of precinct results containing no spatial information. We find asymmetries where Democrats win fewer precincts than Republicans but do so with large marjorities. This skew accounts for persistent Republican control of State Legislatures and Congressional seats even in some states with statewide vote majorities for Democrats.

Despite overall results showing Republican advantages in many states based on mean results from simulations covering many random scenarios, the simulations yield a wide range in metrics, suggesting bias can be minimized better by selecting districting plans with low values for efficiency gap than by selecting plans with values near the means for the ensemble of random simulations.

We examine constraints on county splits to achieve higher compactness and investigate policies requiring cohesiveness for communities of interest as to screen out the most obvious gerrymanders. Minimizing county splits does not necessarily reduce partisan bias, except for Pennsylvania, where limiting county splits appears to reduce bias.

**Introduction**

United States state legislative and Congressional districts are drawn by state legislatures and in a few cases, by non-partisan commissions. These single member, first past the post districts have outcomes that usually differ from the proportion of votes cast. On the Federal level, the 2012 Congressional elections gave Republicans a 7.6% seat majority despite losing the popular vote by 1.2%, although this gap has narrowed considerably in the most recent two Federal elections. In Pennsylvania, a state that as often as not elects Democratic Governors and Senators, Republicans have held the State Senate for 70 of the last 80 years.

Besides ocassionally yielding the perverse result of a party winning a majority of seats with only a minority of the popular vote, single member districts yield disproportinate results for majorities. In the 2020 Congressional elections, Texas Republicans achieved a 28% seat margin with a vote 9% margin. In California, Democrats achieved a 58.5% seat margin over Republicans with a 32.5% vote margin.

Parties have long used the winner take-all feature of American elections to sculpt districts in their favor. North Carolina, Ohio, Texas, Georgia, Pennsylvania, Wisconsin, Michigan, Florida for Republicans and Maryland for Democrats have been the subject of contentious battles. Technology improvements such as the Maptitude districting package have permitted sophisticated partisan districting based not only on current demographics but on projected future demographics using predictive models[1]. The 2019 Supreme Court docket addressed a citizenship question for the Census whereby districts could be shaped to advantage non-hispanic White voters. Questions continue whether States may base districts on citizen or total inhabitant population, potentially diluting representation for immigrant communities.

Even in the absence of mischief, single member first past the post districts result in supermajorities for a party with a modest but spatially uniform statewide majority. How a party translates votes into seat is a subtle function of its geographical distribution. Like Goldilocks and the three bears, it has to be "just right." If a minority party is too widely dispersed, it will not translate seats into votes efficiently. The Republican party in Massachusetts tends to be under-represented for this reason. Conversely, if a minority party is concentrated in just a few districts with large supermajorities supermajorities, it concedes contention elsewhere.

Disproportionate representation has been a research focus, considering the advantage the majority party takes in single member districts with first past the post elections. Kendall & Stuart (1950) developed a 'cube law' showing how the majority party achieves over-representation compared to linear proportionality. Chen & Rodden (2013) predict disproportionate representation for the majority party using a 'logit curve' while also predicting a "loser's bonus" if the minority party is concentrated enough.

---

[1] https://www.caliper.com/mtredist.htm

Although these generalizations analyze over-representation by the majority party, surprisingly the majority party can be disadvantaged, almost never translating its majority into seats if its population is distributed inefficiently. In Pennsyvlania, Wisconsin and Michigan, Democrats average 50% or more over numerous state election cycles but rarely control the state legislatures. However it has been difficult to untangle bias due to partisan gerrymandering from the natural effect of political geography.

Bishop (2009) tells the story how compared to earlier times when America was more homogeneous, over the last thirty years Americans have congregated into like-minded communities based on religion, ethnicity, income and political preference. Yet there are crucial differences in how Democrats and Republicans congregate. Nationwide partisanship maps show broad swaths of red surrounding tiny blue enclaves – the cities. These overstate Republican dominance because most red swaths are thinly populated. Conversely, surrounding the bluest Democratic enclaves are small pink and purple areas determining the balance of power, intractable by visual impressions alone but rather through quantitative work.

The next level of analysis relates to the relative concentration of Democrats and Republicans. As Democrats pack in cities and some old inner-suburbs, they often concede the rest of the suburbs and exurbs to Republicans. Rodden (2019, figure 6.1, p. 168) shows an asymmetry exists between Republicans and Democrats on their relative proximity. He finds that nationwide, Republicans are 5% more likely to live next to Democrats than vice-versa. In Missouri, Georgia, New York, Maryland and Utah this disparity is 10%. That is, while a majority of Democrats are located in cities among other Democrats, a majority of Republicans live elsewhere more dispersed where their nearest neighbors are either the majority city or minority suburban Democrats, in broad generalization. This suggests that Republicans are distributed more efficiently than Democrats for translating votes into seats.

This paper compares the statistical distribution of voter preferences in individual voting districts to full geographical simulations, determining the political geography of the two US parties over ten States. Our sample covers States where Republicans and Democrats have had majorities, together with battleground States. We determine the partisanship skew, the extent of the "losers' bonus" and the relative advantage of the parties in our sample, apart from any gerrymandering.

Many partisan gerrymanders patched precincts together in geographic contortions rivalling the original "Gerrymander" from the early 19$^{th}$ century, a bat-winged creature snaking across Massachusetts tipping the State Senate for the Democratic-Republicans over the Federalists. Public outcry at these blatant abuses resulted in calls for more compact districts where communities of interest remain intact and splitting counties into multiple districts is limited to the minimum feasible. Meanwhile, authors of partisan districts shifted to more subtle tactics where districts pass the "eye test" but sophisticated computer analysis reveals the bias in their construction, showing that despite the "natural" look of the plans they are still one in a billion outliers compared to an unbiased distribution.

Do more compact districts necessarily result in fair districts? What constitutes a fair districting plan? Is a plan constructed without any partisan considerations resulting in 25% of the seats going to the party achieving a 46% vote share more equitable than a plan where more proportional representation is achieved? We investigate the effect of district compactness by imposing county-splitting constraints in our geographic simulations.

**Data and Methods**

We analyzed election data covering 10 US States, tracking the partisan composition of their political geography. Census bureau Voter Tabulation Districts (VTDs) are a base "building block" for constructing political districts for Federal Congressional and State Legislative representation, containing typically a few hundred voters each. They are not necessarily identical to county voting precincts, requiring cross-tabulation and interpolation when obtaining results directly from county election boards. We obtained pre-screened data from the Tufts University Metric Geometry and Gerrymandering project (mggg.org) and from Davesedistricting.org, a non-profit project diffusing mapping tools to the public.

These data contain election results for a range of statewide races per state and ESRI shapefiles for VTDs. These are geometric objects with coordinates defining often irregular bounds, readable into GIS mapping software such as QGIS, Maptitude, ArcGIS, helpful for physically examining map data directly. In some cases point objects or kinks required healing with vector buffering. Many states contain disjoint domains eg. islands that must be connected for further processing (Michigan's Upper Peninsula, Nantucket & Martha's Vineyard).

The VTD data contained or were joined with census population and other demographic information. Field codes for each VTD entry contain county identifiers for geolocation, usually with identifiers for Congressional and State Legislative district assignments for particular political district plans. Where the VTD data did not contain political district boundaries, we obtained them from the Census Bureau and interpolated VTD information into them using the MGGG map interpretation toolkit[2].

We used state-wide election results to infer partisanship tallies per individual VTD not district election results for local candidates. This avoids unopposed races and bias associated with name recognition and constituent services for incumbents, especially in down-ballot races. Actual votes for candidates are more direct indicators of partisanship than party registration, as "Reagan Democrats" in Western Pennsylvania and elsewhere maintained their historical party while voting Republican. Possibly the best partisanship measure is results for low-profile statewide offices like Treasurer or Auditor where people vote their party preference, not based on candidate personality.

Where possible we used composites of multiple elections as opposed to relying on a single election, leavening swings from singularly popular candidates whose charisma motivates voters to cross partisan lines or whose opponents are particularly weak. This is effect is surprisingly large. Examples from Pennsylvania show in 2018, Senator Casey [D] won re-election with a 9.3% party margin while in 2016 Senator Toomey [R] won with a -1.6% party margin, not counting a significant Libertarian showing. Meanwhile Presidential race margins in Pennsylvania have been fairly narrow. A convention for this paper is, with the Democratic Party as the oldest party, Democratic margins will be positive numbers and Republican margins will be negative numbers, applicable to metrics such as [median– mean] and efficiency gap. The list of elections used per state is in Table I.

---

[2] https://www.census.gov/cgi-bin/geo/shapefiles/index.php; https://github.com/mggg/maup

Table 1. Election data used to form composite measure of partisanship, by state

| State | President | Senate | State Offices |
| --- | --- | --- | --- |
| MA | 2012, 2016 | 2012, 2013, 2014, 2018 | Gov 2014, 2018 |
| TX | 2012, 2016 | 2012, 2014 | Gov 2014 |
| FL | 2016 | 2015 | |
| MI | 2016 | | |
| WI | 2012, 2016 | 2012, 2016, | Gov 2012, 2014; AG 2012, 2016 |
| GA | 2016 | 2016 | |
| OH | 2016 | 2016 | |
| PA | 2012, 2016 | 2010, 2012, 2016 | Gov 2010, 2014; AG 2012, 2016 |
| NC | 2012, 2016 | 2014, 2016 | Gov 2012, 2016 |
| MD | 2012, 2016 | 2012, 2016, 2018 | Gov 2014, 2018; AG 2014, 2018 |

We used the Gerrychain software package produced by MGGG[3] to partition random assortments of VTDs into districts using Markov chains. We used the recombination (RECOM) algorithm at each Markov step described in detail elsewhere (Deford, Duchin & Solomon 2019) so we summarize briefly. Starting with an initial seed, which may be either an actual enacted plan or for cases where these were not well-behaved, a random seed, a subsequent Markov step is proposed by combining two randomly selected adjacent districts then slicing the combined district along a randomly selected boundary between VTD's. The proposed Markov step is accepted if the partition as a whole meets constraint criteria, eg. on population deviation per district.

Responding to efforts to reduce gerrymandering by requiring compactness and limiting how political entities like counties may be parceled among numerous districts, we included compactness constraints and limits on county splits in some model runs. To constrain compactness we excluded proposals whose boundary length exceeded double the intial state. To limit county splits, first we tallied the number of counties allocated to multiple districts in a proposed chain step. If this tally equaled or was below the maximum permitted number of county splits we accepted the proposal for the Markov chain. If the split tally exceeded the maximum permitted number of splits but did not exceed the tally from the prior Markov step, we retained the proposed step for step evolution but did not evaluate the new Markov state in the statistical analysis for redistricting nor did we count the step towards the maximum step limit for the chain. In this way, a chain is forced to evolve, eventually, to a state with the required maximum number of county splits. Checks indicated runs witth a random initial state having many county splits and a prepared initial state with relatively few splits had similar statistics. Thus where a usable initial state with a low number of county splits existed, we used it.

---

[3] https://github.com/mggg/GerryChain

The process is computationally intensive. We used a 40 core server with 4x Intel E7-4870 CPUs at 2.4GHz, having optimized gerrychain for multi-threading computation (40 parallel threads) accelerating roughly 35x compared to single core computation. The random number generator for each of the 40 Markov chains was initiated with a different seed while each chain was initiated with the same districting partition, ensuring independent chain evolution. Each was extended for 500 steps, thus reaching 20,000 states. Without county split constraints, this many states could be reached in a few minutes. With split contraints, this number of states could take hours or overnight to run, if the chain would converge at all.

Each step in a Markov chain is not independent of the previous step. Like cutting a card deck only once, a single step only shuffles a Markov chain somewhat. We found chain correlation to the initial step decayed to 0.2 in roughly 50 steps depending on the State and district type, fluctuating thereafter with lower corrrelation. However, some State House races never became uncorrelated, possibly trapped in a local space by population or other constraints, with few VTD's composing each district providing fewer randomization options, sometimes becoming poorly behaved and not converging ever even without county split constraints. Thus for this study we only considered Federal Congressional and State Senate races with their larger districts.

**Results and Discussion**

We characterized the VTDs by creating histograms of the partisanship distribution for the datasets covering the ten states in our sample. For each district a VTD represents we tabulated its partisanship within 1% bins, weighing the VTD count in that bin by the mean voting population per VTD. This differs from slightly weighing each VTD by its *number of inhabitants* because we account for voting turnout. This differs as well from a conventional histogram where each bin contains a simple VTD count of the # VTDs within a partisanship. These histograms are shown in Figure 1. The histograms show the cumulative total fraction of voting population on a secondary vertical axis. Being voting population equal-weighted, these histograms are comparable to actual geographical simulations, unlike histograms counting towns or other unequal entities.

Thus, these histograms tally the partisan voting power on a district basis, showing the proportion of equal voting districts at different percentiles of voting strength. Naively one might expect the share of elected Democratic representatives equal the proportion of VTDs with Democratic majorities. Another indicator is the partisanship % at the 50% voting population point. If if differs greatly from 50% we can expect a strongly partisan outcome for the allocation of seats. A key question: will a party with a minority of VTDs capture an equal proportion of seats? Shedding light on the translation of votes into seats is the primary investigation of this paper.

Figure 1. Precinct partisanship distribution computing Voter Tabulation District histograms, showing # equal vote-weighted VTD's within each 1% bin of Democratic vote share % (black), cumulative voting population % (blue). Green vertical line at 50% voting population vote share. Blue vertical line at mean Democratic vote share %. Black vertical line at 50% Democratic vote share point for reference. Where the green line is to the right of the black line, the median precinct has a Democratic majority, and vice-versa. Black vertical line at 50% Democratic vote share point for reference. a. Massachusetts; b. Texas; c. Florida; d. Michigan; e. Wisconsin; f. Georgia; g. Ohio; h Pennsylvania; i. North Carolina; j. Maryland

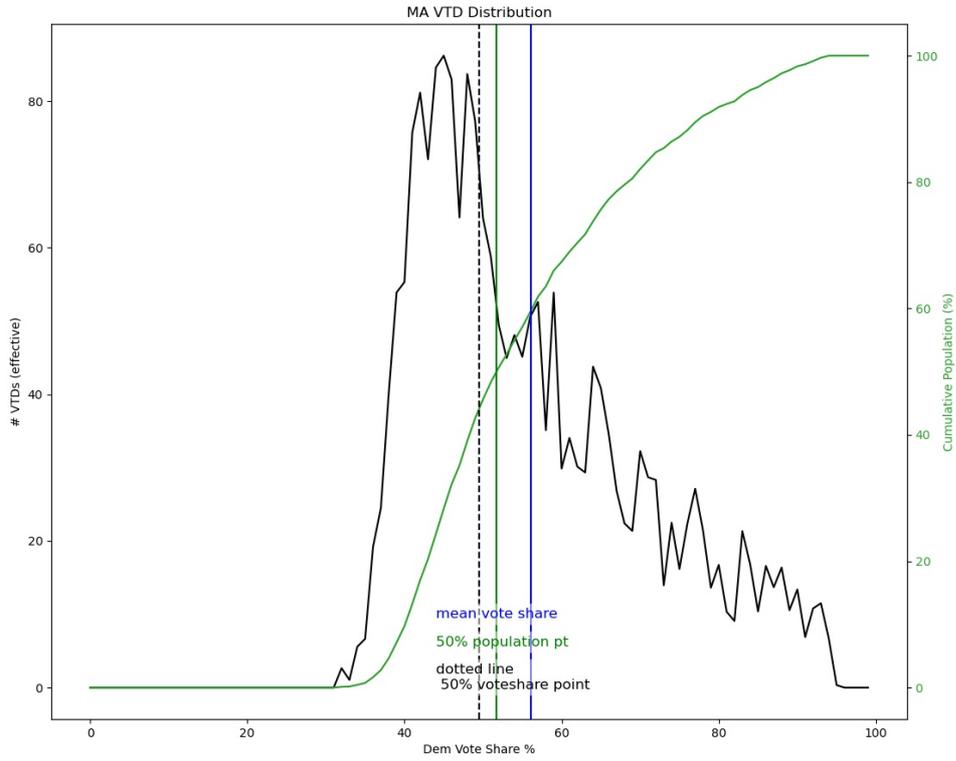

a. MA

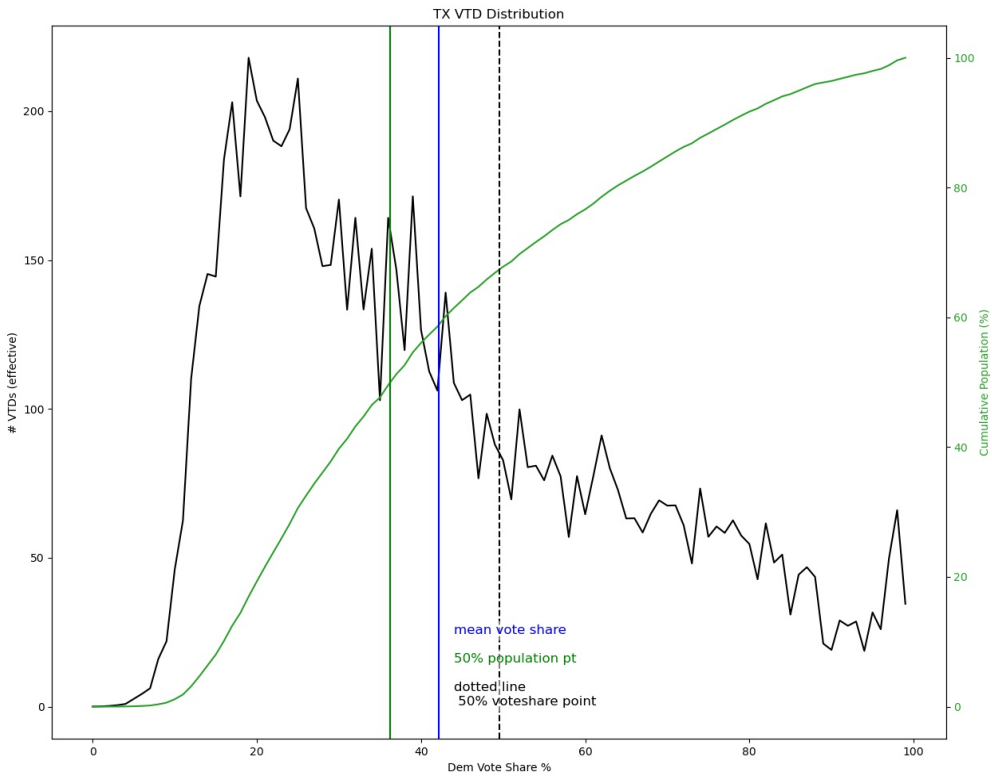

b. TX

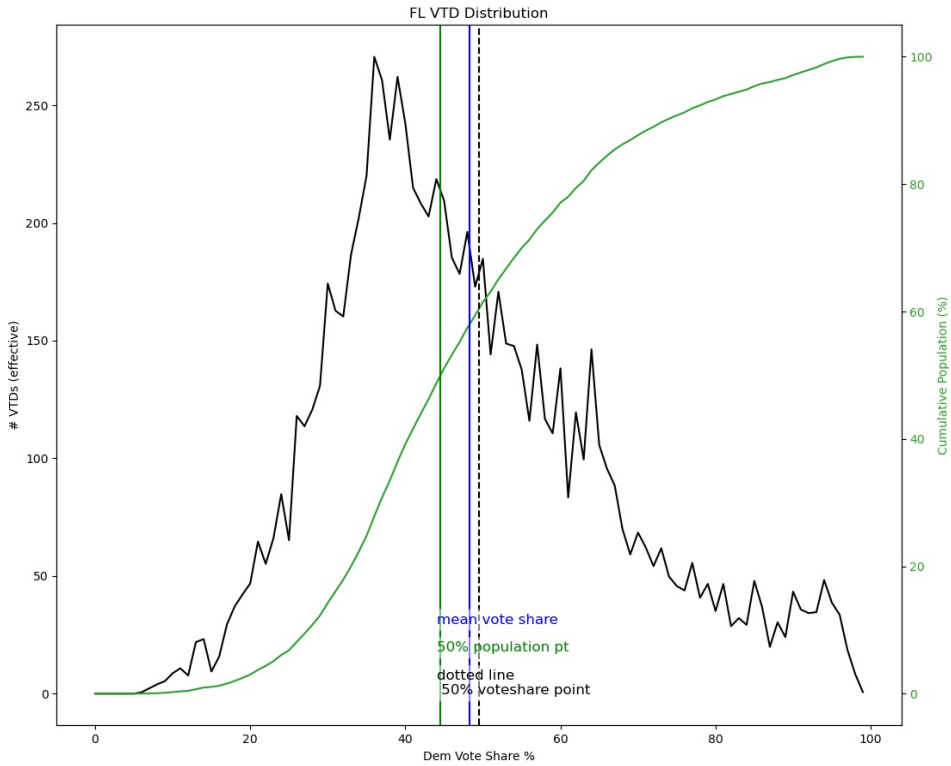

c. FL

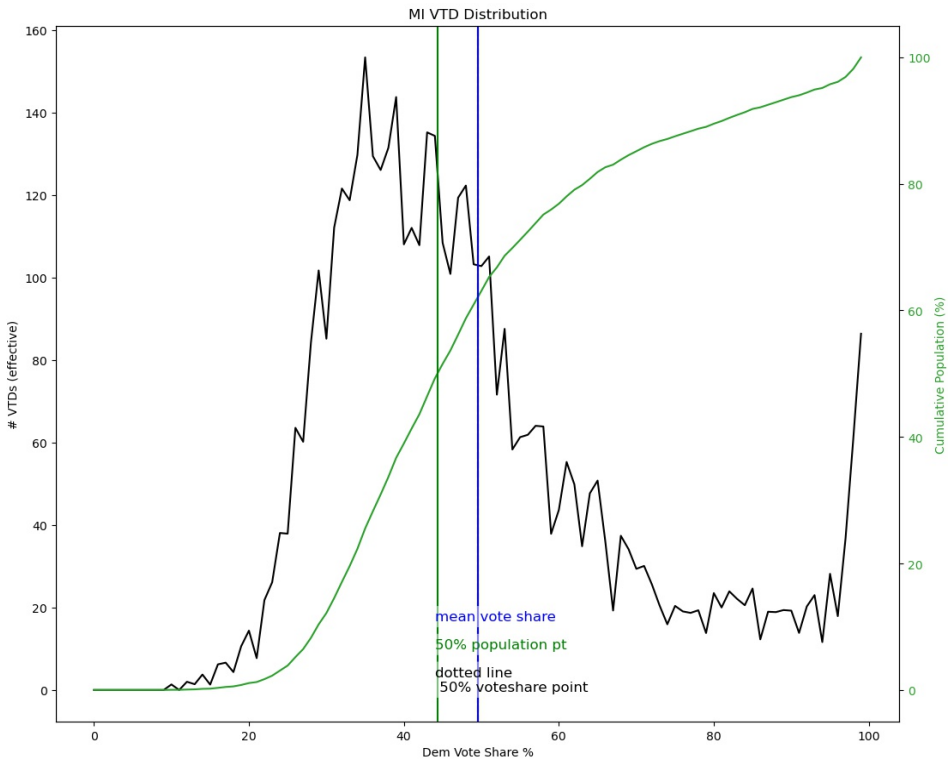

d. MI

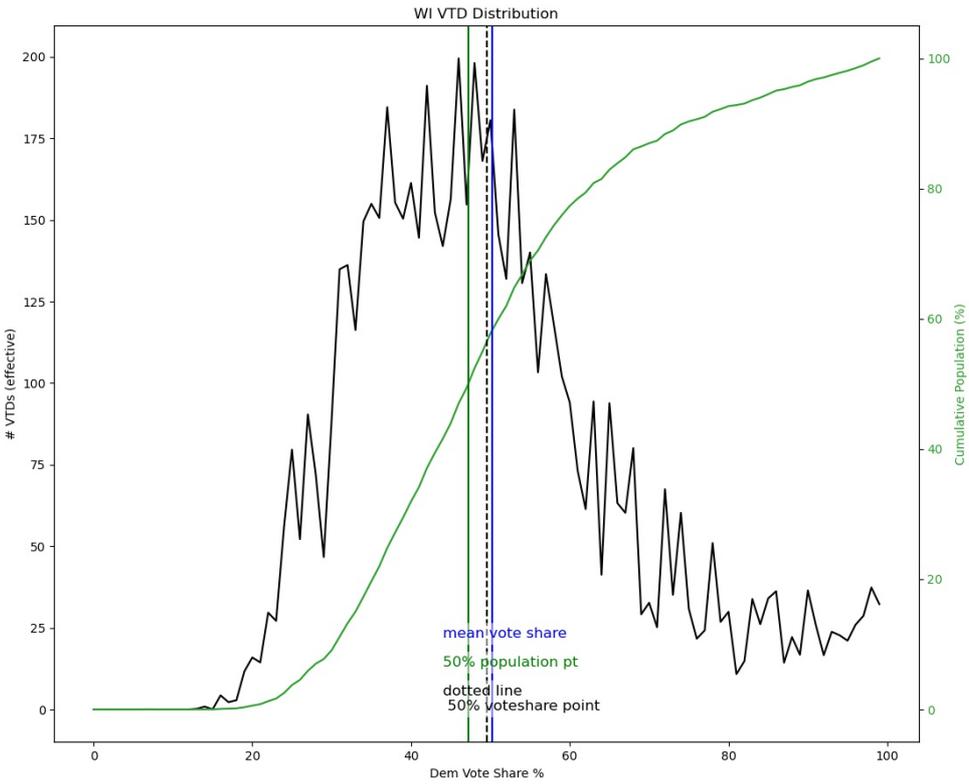

e. WI

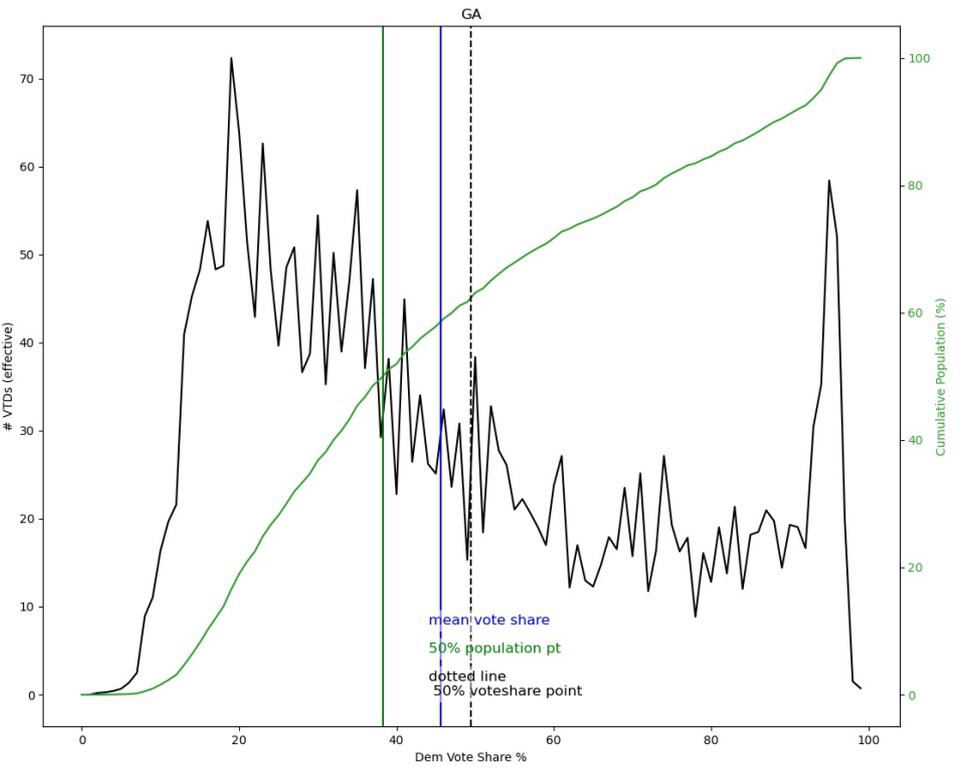

f. GA

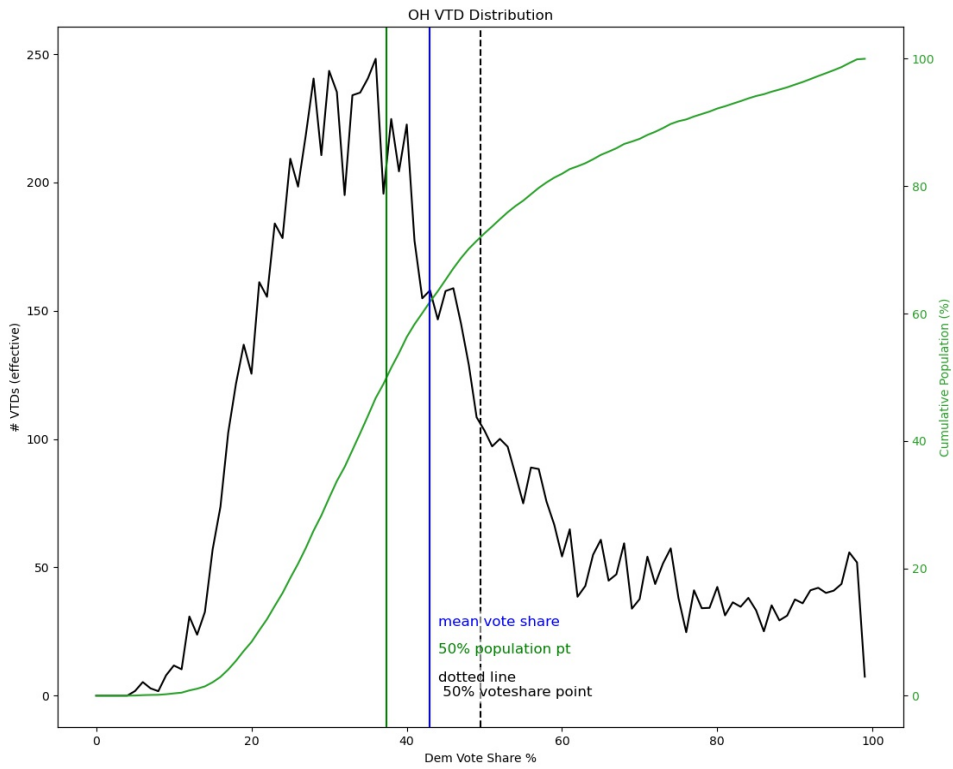

g. OH

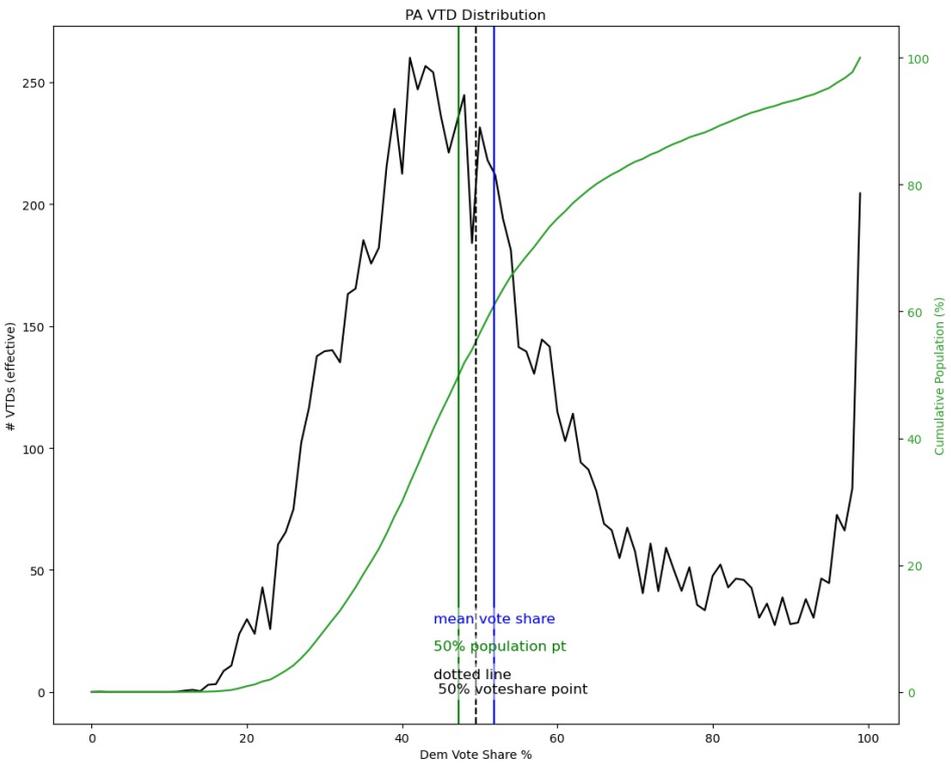

h. PA

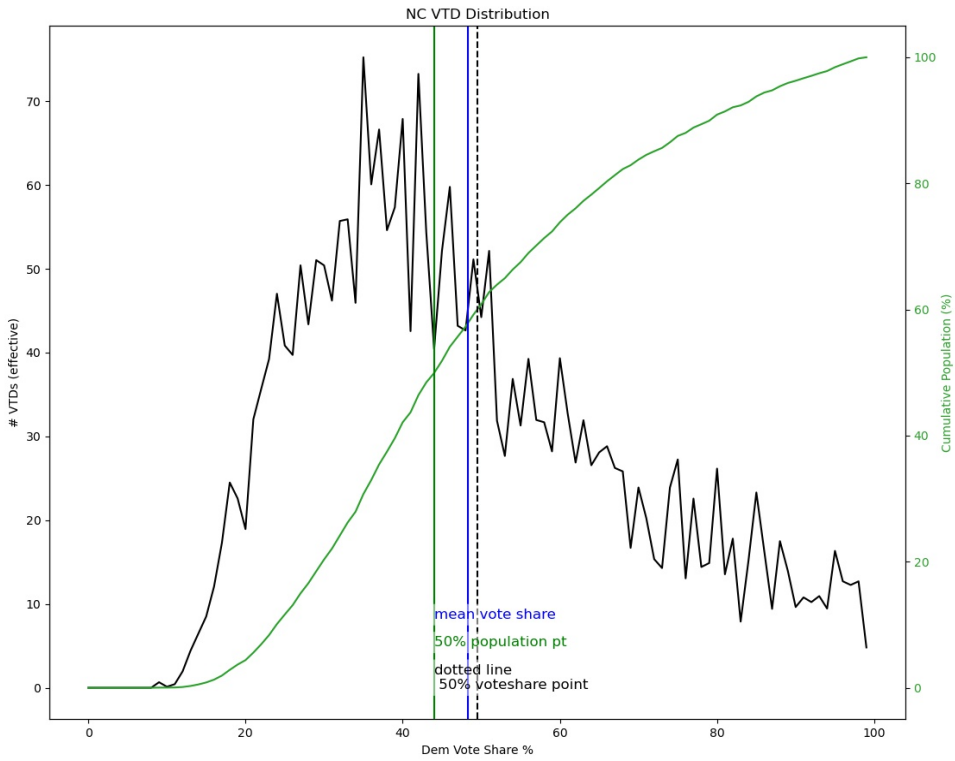

i. NC

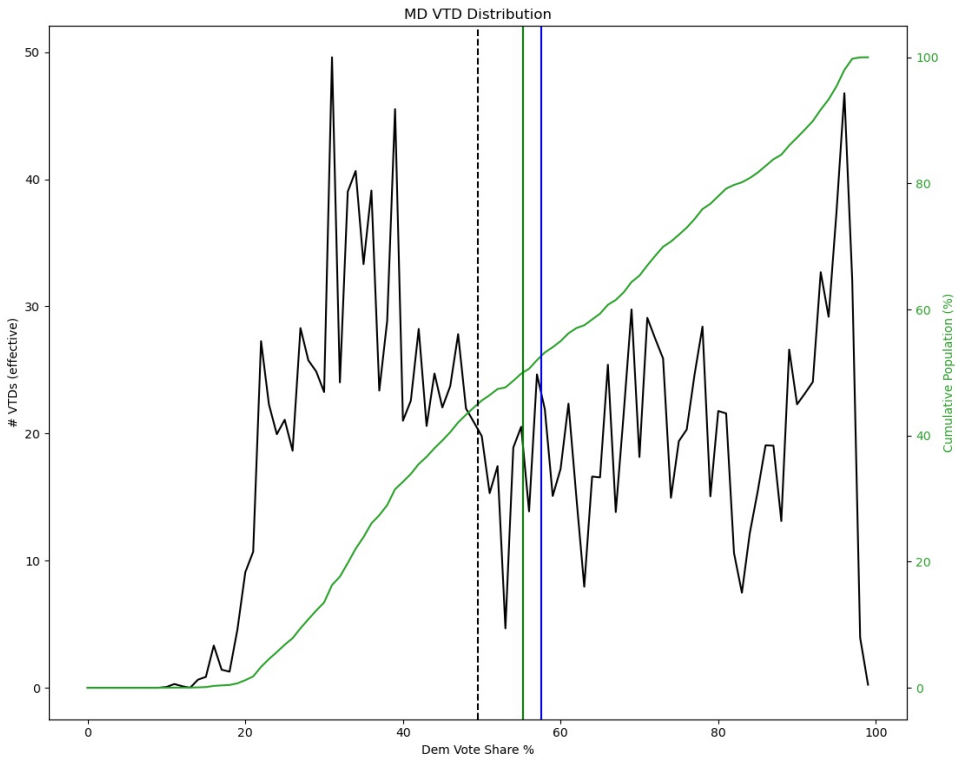

j. MD

Figure 1 shows the striking result that every state shows a high degree of skew in the partisanship distributions, often with sharp peaks in numbers of VTDs with extreme Democratic vote margins over 80%, eg MD, MI, GA, PA. States with asymmetric slopes rightward in partisanship distribution are MA, FL, TX, NC, OH. Although we don't present detailed data here, partisanship color maps of states display tiny knots of Democratic strongholds in cities, neutral zones around cities like a halo around the moon on a humid night, with broad swaths of Republican territory everywhere else. The rural/ urban split partisan split is detailed further in Rodden (2019).

Democratic concentration facilitates gerrymandering. Classically, biased districts snake across a state linking separate urban areas together with corridors as narrow as a lane on an interstate highway, as was done after the 2010 apportionment in North Carolina. But legislators can gerrymander districts without 'batwings' using more subtlety (Duchin 2018) by pairing heavily Democratic cities with the most Democratic inner suburbs while giving a partisan edge to mid-range suburbs.

Heavily skewed geographic distributions like these advantage Republicans even without any gerrymandering for two reasons: Democratic precincts in cities are more overwhelmingly Democratic than rural precincts are Republican, and urban precincts cluster together inefficiently. Votes don't translate into seats efficiently when large number of hyper-partisan precincts are adjacent and can not be paired easily with less partisan districts due to geography.

While a political minority only captures seats if there is *some* clustering, the first past the post election system also create inefficient representation if there is *too much*: in this sample, Democrats received 52.4% of the votes in Pennsylvania but have been in the minority in the State Senate for 70 of the last 80 years, a stunning failure to translate votes into seats because of democratic vote clustering in Philadelphia and Pittsburgh. We begin our analysis of partisan representation with the VTD distribution statistics and each party's mean vote share in Table 2.

Table 2 shows the base voter statistics with Democratic states (MA & MD, mean vote shares 0.566 and 0.581), Republican states (TX & OH, mean vote shares 0.427, 0.434) and other states in between. Every distribution has positive skew except for MD, which appears nearly bi-modally symmetric. The standard deviation in vote share is as low as 0.143 (MA), an outlier, and as high as 0.259 (GA). We introduce new measures: the Democratic vote share for the median VTD, and the Democratic total % of VTDs at the 50% vote share VTD. If the vote share at the median VTD < 50% that party will capture a minority of the VTD's even if that party captures a majority of the statewide votes. The VTD% share at 50% vote share is an indicator how many seats a party may capture *if* VTD majorities translate to seats.

Table 2. Precinct statistics for precinct partisanship distribution over composite measure of all sampled election results. Mean Democratic vote share, standard deviation, skew, kurtosis, Democratic vote share for median precinct, Democratic population share at > 50% vote share.

| state | mean vote share | stdev vote share | skew | kurtosis | Median VTD Dem vote share | Dem VTD share at >50% vote share |
|---|---|---|---|---|---|---|
| MA | 0.566 | 0.143 | 0.777 | 2.713 | 0.522 | 0.559 |
| TX | 0.427 | 0.231 | 0.675 | 2.454 | 0.368 | 0.327 |
| FL | 0.487 | 0.180 | 0.642 | 3.015 | 0.450 | 0.396 |
| MI | 0.500 | 0.194 | 0.980 | 3.286 | 0.448 | 0.380 |
| WI | 0.508 | 0.172 | 0.808 | 3.308 | 0.477 | 0.436 |
| GA | 0.462 | 0.259 | 0.562 | 2.080 | 0.387 | 0.376 |
| OH | 0.434 | 0.203 | 0.973 | 3.285 | 0.379 | 0.280 |
| PA | 0.524 | 0.191 | 0.859 | 3.130 | 0.478 | 0.448 |
| NC | 0.488 | 0.202 | 0.593 | 2.560 | 0.445 | 0.400 |
| MD | 0.581 | 0.237 | 0.161 | 1.705 | 0.557 | 0.550 |

Several investigators have suggested relationships between the number of seats 'S' and votes 'V' earned by a party, starting with the 'cube law' of Kendall & Stuart (1950) whereby $S/(1 - S) = (V/[1-V])^3$. More recently Chen & Rodden (2013) used a 'logit model' to predict seat share by comparing vote results to an inflection curve. Here we build intuition from a naïve two-factor model just to estimate the portion of VTD's won by a minority party given a quasi-gaussian distribution characterized by its first and second moments:

$$f(x) = \frac{\exp -\pi (x - \mu)^2/\sigma}{\sigma\sqrt{2\pi}}$$

With a VTD share approximated by

$$VTD_{share} = \frac{\int_{0.5}^{1} f(x)dx}{\int_{0}^{1} f(x)dx}$$

This is not quite a gaussian error function or cumulative probability density with standard deviation σ and first moment μ because *f(x)* has non-zero values < 0 and > 1 although the range 0 < x < 1 captures most of the probability density for actual values of σ and μ. The denominator is a normalization factor. The key feature for our intuition is for a minority party, with low standard deviation σ, very few VTDs are captured but as σ increases, the fraction of VTDs capture asymptotically approaches vote share.

Figure 2: Naïve 2-factor model for precincts captured by minority party with 45% vote share as function of standard distribution of vote standard deviation, σ.

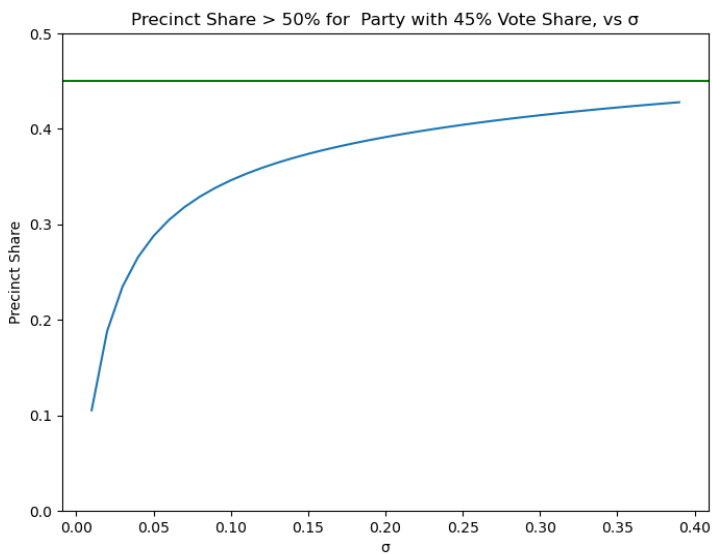

For our dataset we can do better because we possess the actual distributions *f(x)* – they are simply the histograms from Figure 1. In Table 3 we present the results from the Markov chain model and in Figure 3 we compare the Markov chain result to the simple VTD seat share model. The chart shows surprisingly good agreement between the VTD share model – which incorporates no geographical or density correlations – and the full geographic analysis afforded by the Markov simulations.

We expect larger districts - Congressional seats as compared to State Senate districts - will have larger variance in voter density as spatial fluctuations aggregate randomly with a Poisson-like distribution. Thus they would have greater deviations from the simple VTD share model assuming uniform spatial density. This is so, including Massachusetts where the simulations deviate from VTD share the most (Figure 3).

Table 3 presents the Markov chain results for [median– mean], efficiency gap and predicted seat share for Congressional and State Senate seats. Here, the number of county splits was the lowest that would converge. Thus the districts were more compact than other random configurations less constrained, thus more closely resembling actual districts hand-drawn to maintain communities of interest. With the lowest skew of the 10-state sample, Maryland has neutral metrics. Other than neutral Maryland and heavily Democratic Massachusetts, where [median– mean] favors Republicans but efficiency gap favors Democrats, all the metrics favor Republicans. That is, even without gerrymandering, for cases with low county splits, the political geography is such that Democrats face hurdles translating votes into seats. The metrics favor Republicans follows because of skew and clustering of low efficiency VTD's with overwhelming Democratic majorities. For example, we find VTD skew and [median– mean] have a -.6 correlation coefficient for State Senate seats with a somewhat lower value of -.41 for Congressional seats, where higher spatial variance introduces noise.

Massachusetts is anomalous because for over 20 years it has sent no Republicans to Congress and it currently only has 3 Republican State Senators despite a long history of Republican governors (Weld, Cellucci, Romney, Baker) - only one Democratic Governor since 1991. Duchin et al. (2019) argue Massachusetts favors Democrats because Republicans are too geographically dispersed, thus do not agglomerate enough VTD's to carry whole districts, and obtain zero Republican Congressional districts in their simulations. Notably, with an anomalously low standard deviation in the vote distribution of the 10-state sample, Massachusetts is the least likely to give the minority party representation commensurate with vote share.

Efficiency gap is a measure not only of geographical advantage, it's a measure of the overall vote share considering for a party with a 40% share everywhere, all of their votes will be wasted while for the party with 60% share, only 10% of their votes will be wasted. Conversely, [median– mean] reflects only distributional characteristics. Thus the highly positive value for efficiency gap in Massachusetts is not determinative of geographical problems for Republicans considering the negative values for [median– mean] reflecting distributional differences without bias from mean vote share.

The simulations show that on average, Republicans might expect to win 1.8 Congressional seats based on state-wide vote totals and 10.0 State Senate seats, thus differing with the Duchin et al. results. Our dataset may differ from Duchin's as it includes election results such as the 2018 gubernatorial race that Republican Charlie Baker won overwhelmingly in 2018 (64% - 36%).

Table 3.

Mean statistics for Congressional & State Senate Seats and Predicted Seats from Markov Chain simulations together with [median– mean] and efficiency gap. Statistics are mean values from 400 independent scenarios. Positive numbers for [median– mean] and Efficiency Gap favor Democrats.

| State | Mean vote share % | Con. Seats | Pred Congress Dem Seats % | Senate Seats | Pred Dem St Senate Seats % | Median – mean Congress | Efficiency gap Congress | Median-Mean St Senate | Efficiency Gap St Senate |
|---|---|---|---|---|---|---|---|---|---|
| MA | 56.6 | 9 | 80.2% | 40 | 75.1% | -0.0287 | 0.095 | -0.0275 | 0.05813 |
| TX | 42.7 | 36 | 31.5% | 31 | 32.5% | -0.0141 | -0.0523 | -0.0187 | -0.04242 |
| FL | 48.7 | 27 | 39.9% | 40 | 43.1% | -0.02395 | -0.0848 | -0.0121 | -0.05263 |
| MI | 50.0 | 14 | 39.1% | 38 | 40.8% | -0.02775 | -0.1096 | -0.0509 | -0.09525 |
| WI | 50.8 | 8 | 34.9% | 33 | 41.2% | -0.0361 | -0.1402 | -0.0255 | -0.0868 |
| GA | 46.2 | 14 | 30.4% | 56 | 35.7% | -0.045 | -0.089 | -0.0517 | -0.04624 |
| OH | 43.4 | 16 | 24.1% | 33 | 27.3% | -0.005 | -0.10506 | -0.0288 | -0.0764 |

| PA | 52.4 | 18 | 47.6% | 50 | 45.9% | -0.0185 | -0.037 | -0.0285 | -0.0621 |
| NC | 48.8 | 13 | 34.4% | 50 | 41.0% | -0.0183 | -0.08838 | -0.0225 | -0.036 |
| MD | 58.1 | 8 | 65.3% | 47 | 59.4% | 0.012 | -0.0076 | 0.00185 | -0.0761 |

Figure 3: Comparison of Democratic VTD Share % (actual historical voting data, blue columns) to simulated seat share % for Congressional Seats (red columns) and simulated share for State Senate (green columns)

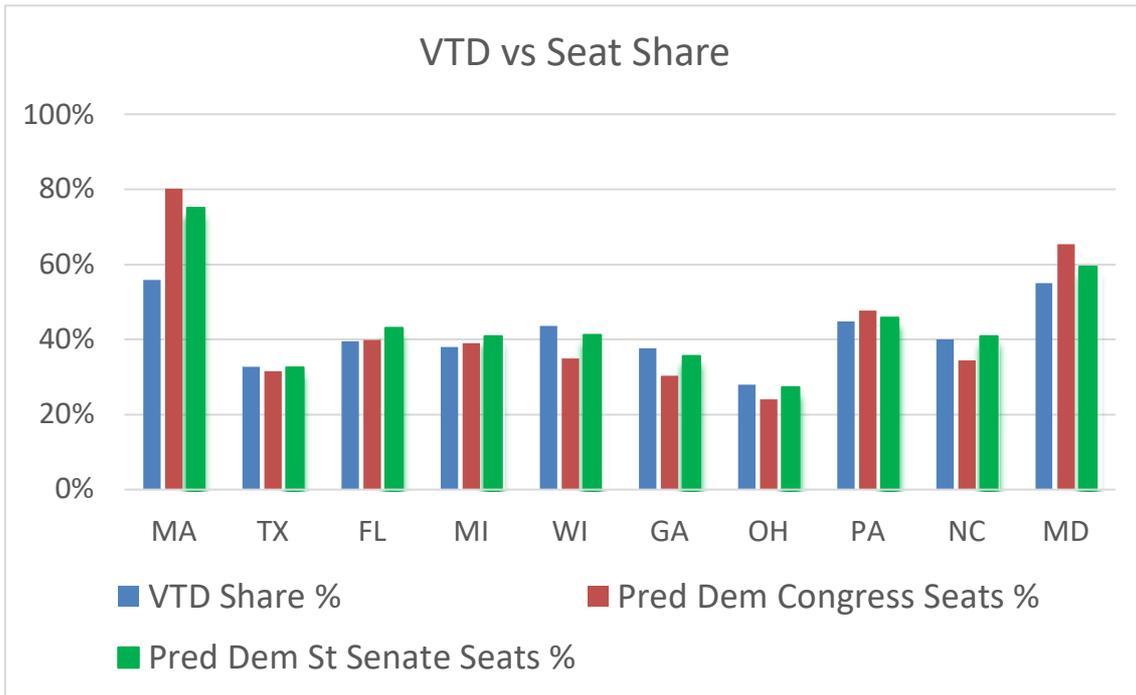

Figure 3: Comparison of Democratic VTD and Seat Share

The above statistics represent mean values (eg. for efficiency gap, median – mean], providing an indication for political geography. Table 3, with large negative values for efficiency gap and [median– mean] for all states except Maryland and possibly Massachusetts shows the structural disadvantage Democrats face in those states, and conversely, the disadvantage Republicans face in Massachusetts.

The mean simulation results provide an indication on the political geography for the states in our sample. A conventional analysis to elucidate gerrymandering's impact is looking for differences between these mean values and their measures in enacted districts. One result of our investigation on political geography is that mean values for seat share can still depart considerably from vote share. Big differences between mean vote share and predicted seat share exist (Table 3). Apart from the mean values of metrics in the simulations there is a large range in values, with [median – mean] ranges from the box and whisker plot shown in Figure 4. In some cases, district plans restricting the number of county splits show narrower ranges of this metric than district plans with county splits unconstrained.

A "blind" redistricting selects a random configuration with metrics close to mean for an ensemble, often characterized as "fair." At least if not more "fair" is a districting with an efficiency gap close to zero, as was mandated by Missouri law until a ballot question in 2020 overturned this requirement. Notably, even in the most partisan states such as MA, TX or OH, district plans with near [median – mean] are possible.

Figure 4: Box & whiskers plot of [median – mean] distribution for congressional & state senate seats in simulated scenarios. Boxes represent middle two quartiles (25% and 75% percentile points). Whiskers represent the extent of highest outliers within 1.5 x (Q3 – Q1) beyond middle two quartiles, circles represent further outliers[4]. Distributions for a. no county split contraints; b. minimum county splits obtained. For these runs, maximum population deviation was 10% for state senate districts, 3% for congressional districts.

a. County Splits Unconstrained

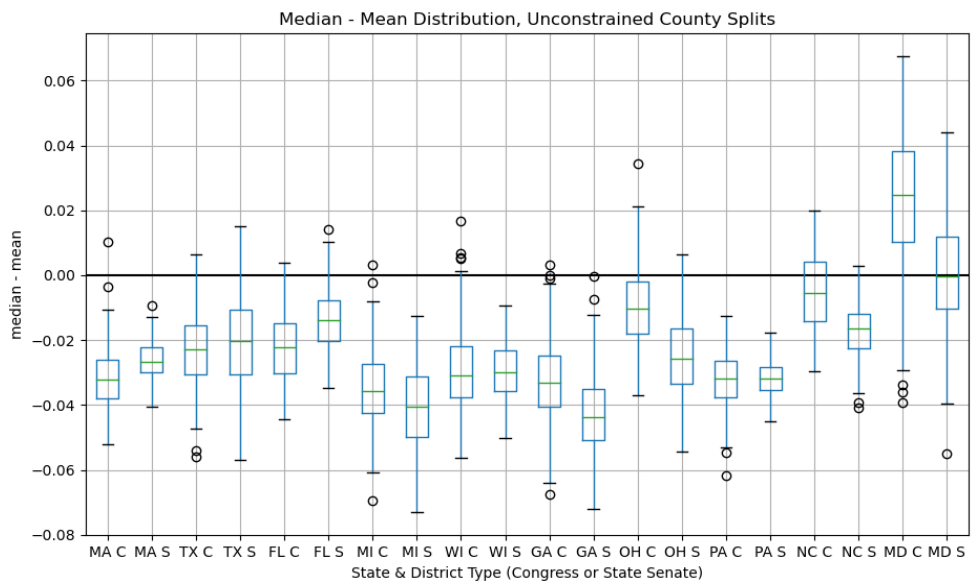

---

[4] See matplotlib documentation for boxplot:

https://matplotlib.org/3.2.2/api/_as_gen/matplotlib.pyplot.boxplot.html

b. Minimum County Splits

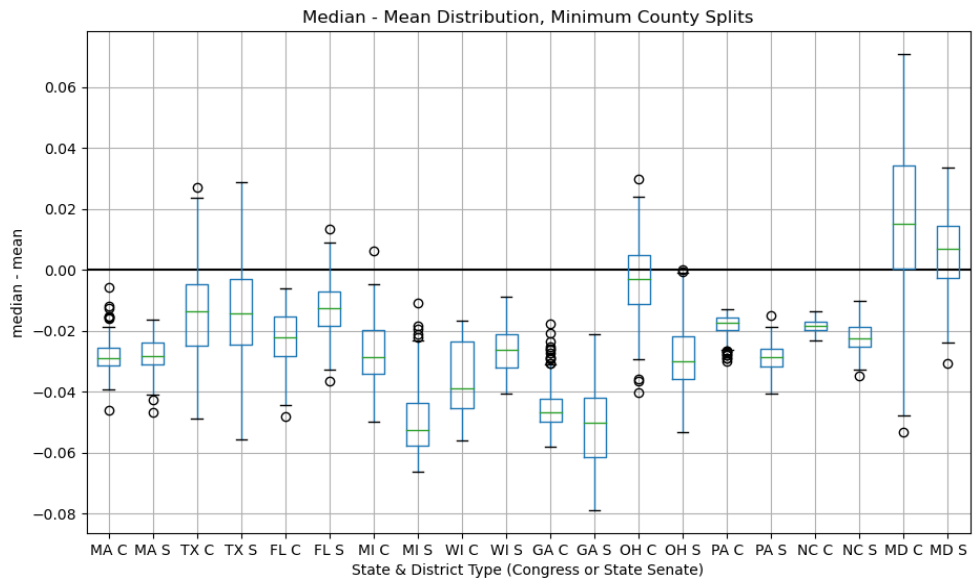

Our final analysis examines how varying the number of county splits affects seats won and metrics. Compactness, here indicated by county splits is often a fairness criteria, especially since notable gerrymanders have entailed districts snaking across many counties to loop a few precincts together here and there for optimal "packing." Several reform proposals place strict limits on compactness and county splitting so assessing how effective these limits are for is needed instead of assuming compactness and preserving community cohesiveness results in better matching between vote share and seat share. Examples of proposals and enacted rules are PA HB22, placing explicit limits on county splits, Ohio's Constitutional requirements for districts, requirements for county integrity in Colorado and Arizona among others.

In Figure 5 we show two-axis plots for each state and for Congressional and State Senate seats of [median– mean] [left vertical axis, in magenta], seats won [right vertical axis, blue] as a function of county splits. We show [median– mean] rather than efficiency gap to avoid bias against the minority party in the efficiency gap measure.

Not taking into account the large scatter discussed above, here we plot the mean values of [median – mean] and seats won over samples of 400 random plans for each split limit scenario. The scatter in the mean outcomes over county constraints indicates a complex picture where no hard and fast rule translates compactness into bias reduction.

Focusing first on Pennsylvania, reducing county splits appears to reduce [median– mean] and has an impact on Democratic seats won for both Congressmen (18 districts) and State Senators (50 districts), consistently in the same direction. Reducing splits from 70 to 18 reduces the extent of [median– mean] from -0.03 to -0.017, with a corresponding increase in Democratic Congressional seats from 7.9 to 8.6 seats, though still falling short of the votes won fraction. A similar but smaller trend is visible for the smaller State Senate districts.

This averages over two different geographic portions of Pennsylvania – the West where Democrats are concentrated in Allegheny county (Pittsburgh) and the East where beyond Philadelphia, Democrats are present in many suburban counties together with smaller cities such as Reading, Easton, Allentown. In the West, Democrats could only achieve more efficient seat shares if Allegheny were sliced up like a pizza, with a core from Allegheny county combined with more evenly divided or Republican precincts from collar counties. Conversely, in Eastern PA results depend on which pieces of which counties combine to form districts but not so much on the number of exact splits given the extent of Democratic vote share in populous counties (Montgomery, Chester, Delaware, Bucks) which then get combined with more balanced or Republican counties adjacent (Berks, Lehigh, Lancaster). An example of an evenly balanced plan is the N9 map for Pennsylvania Congress in Nagle (2019), Figure 4, showing approximate partisan balance at the expense of geographical compactness and cohesion of municipal boundaries. Representation for municipalities mirrors issues with the Voting Rights Act: does a community have the most parliamentary power when it wins seats outright yet only achieve a parliamentary minority, or when it has influence as a significant voting block within districts where it may not be the exclusive power yet instead covers a majority of seats?

Similar trends appear for Ohio, where with the fewest county splits [median–mean] is reduced to almost zero for Congressional seats even though the efficiency gap of -0.105 is only exceeded by Wisconsin (-0.140). Michigan also displays reductions in bias with lower county splits for Congressional seats, however the trend is opposite for State Senate. Maryland is also in this category where reduces county splits reduces [median– mean] for Congressional seats but increases it for State Senate. Other states show no clear pattern so it is hard to say that as a rule, fewer county splits reduces bias. This confirms our belief that other things being equal, splitting counties can be equitable if doing so forms districts with more proportional outcomes, subject to reasonable constraints on compactness.

Figure 5: Left hand plots (Congress), Right hand plots (State Senate) seats and [median – median] as a function of county splits in districting simulations. Left axis (magenta) shows mean values for [median – mean] over 400 scenarios; right axis (blue) shows mean values for Democratic seats won over the scenarios.

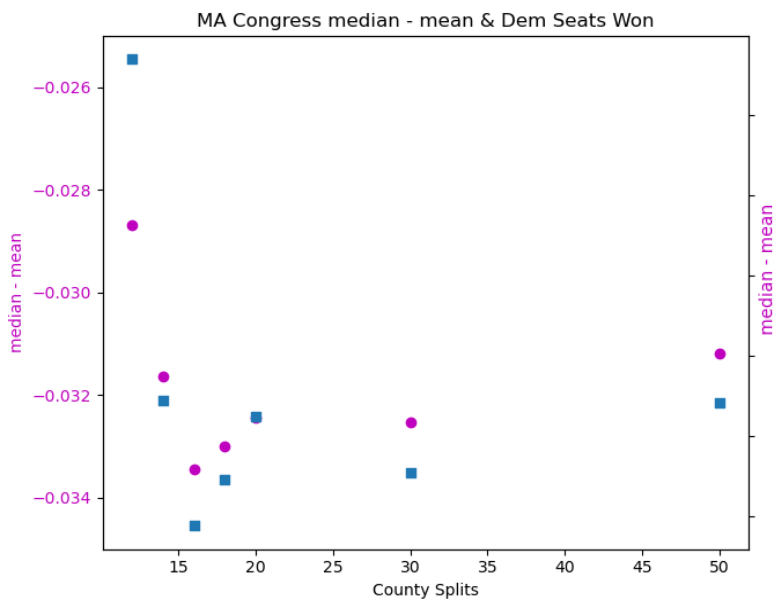
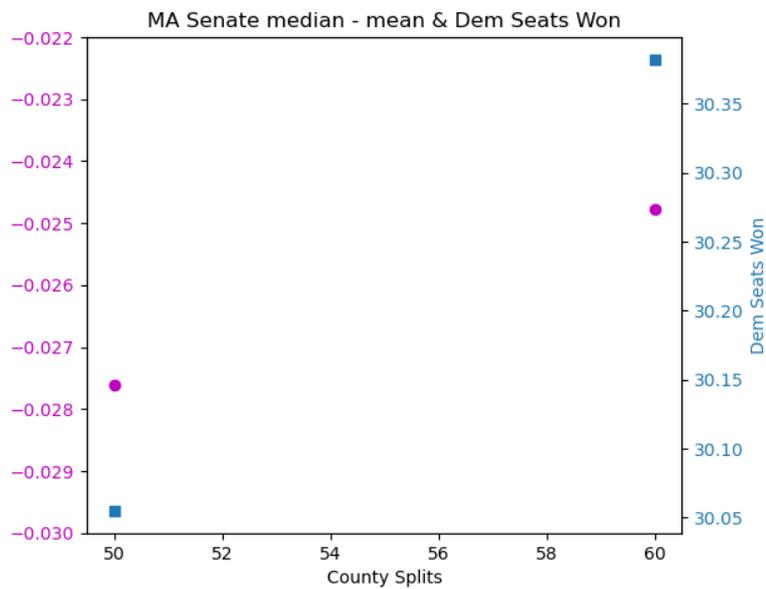
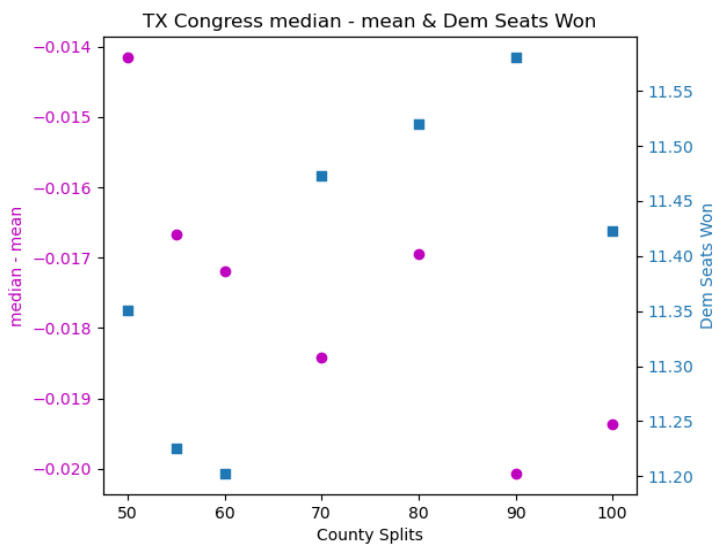
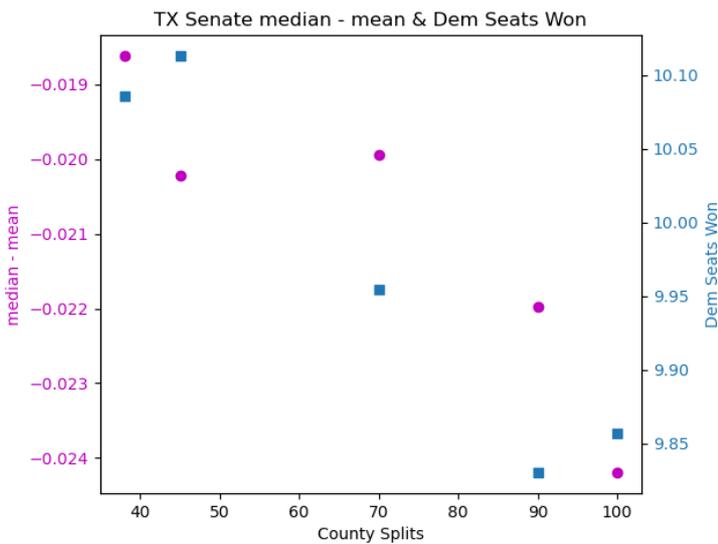

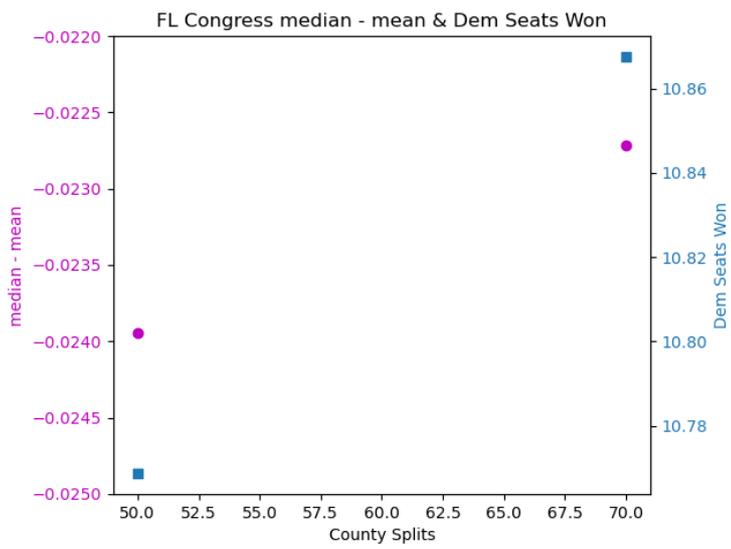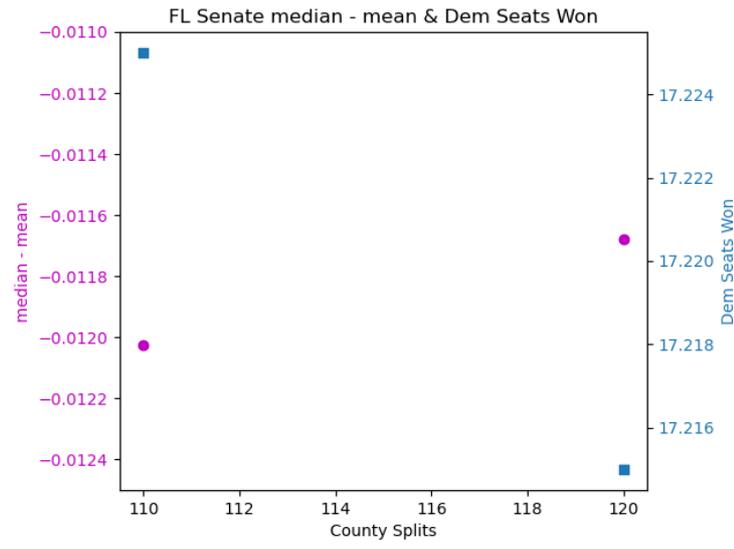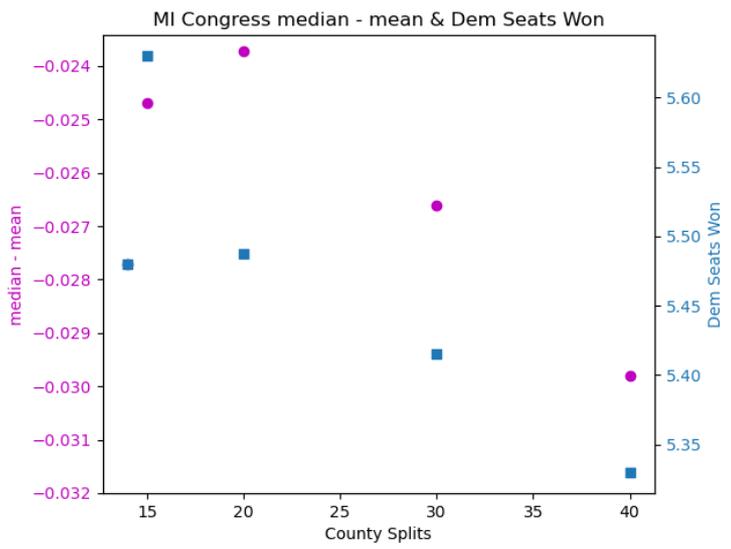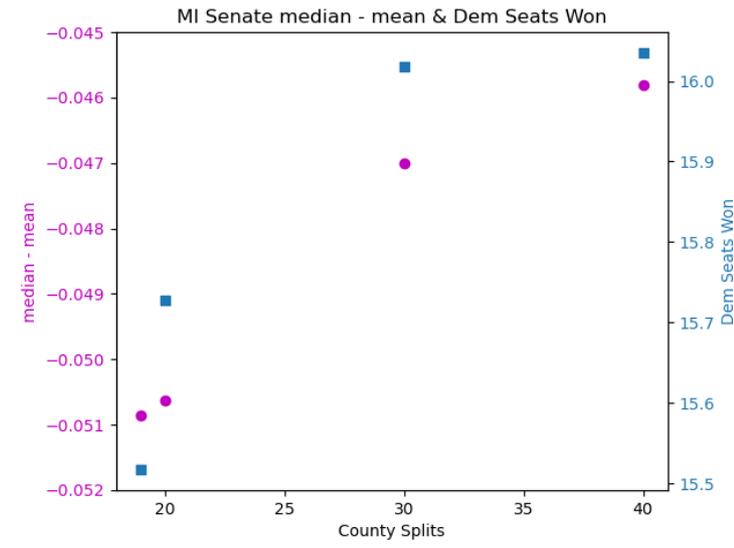

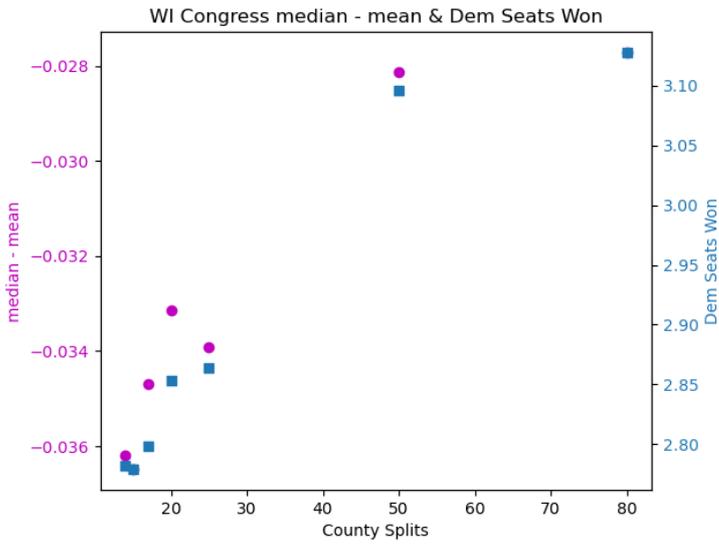
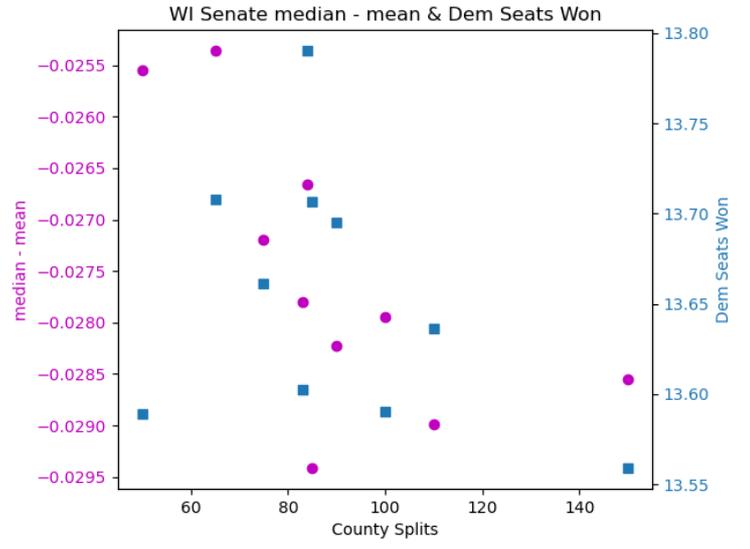
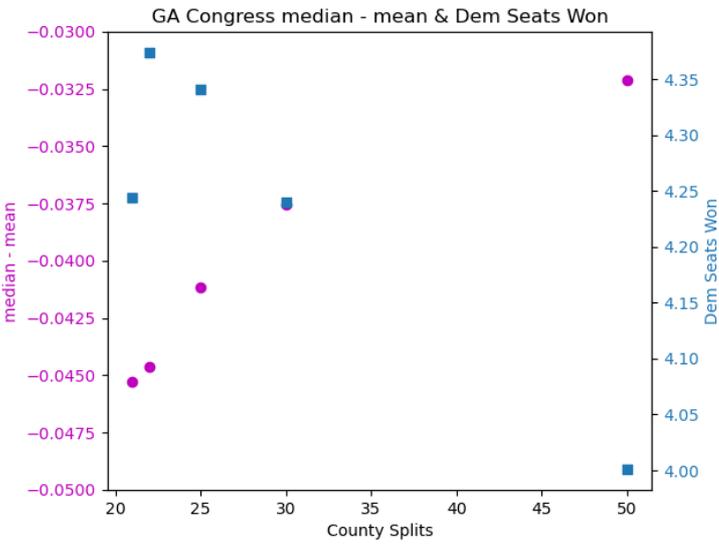
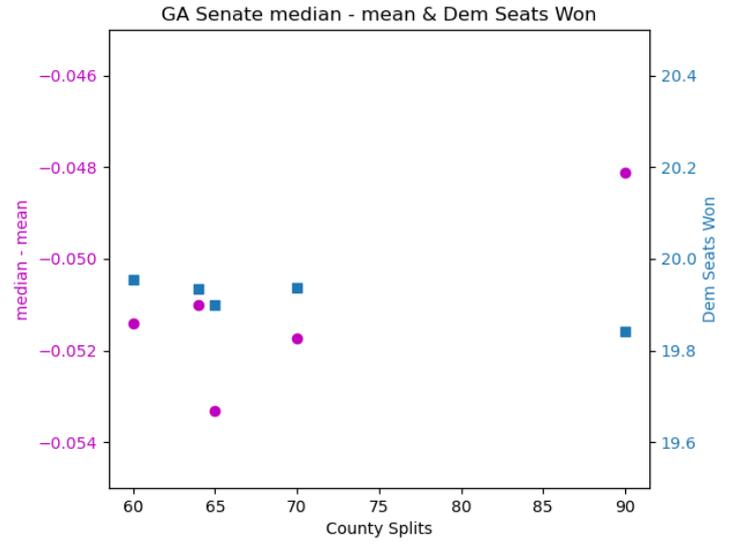

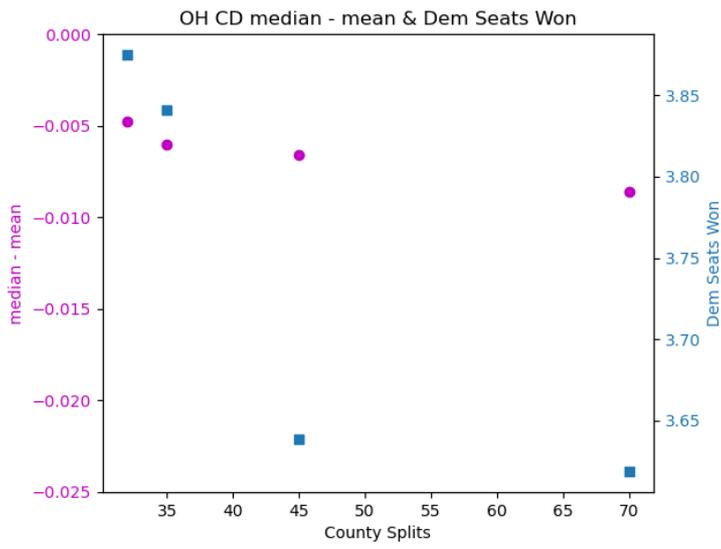
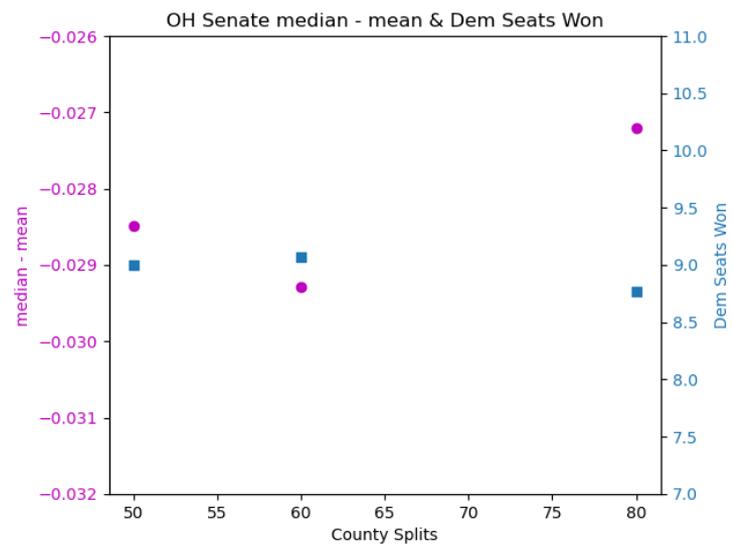
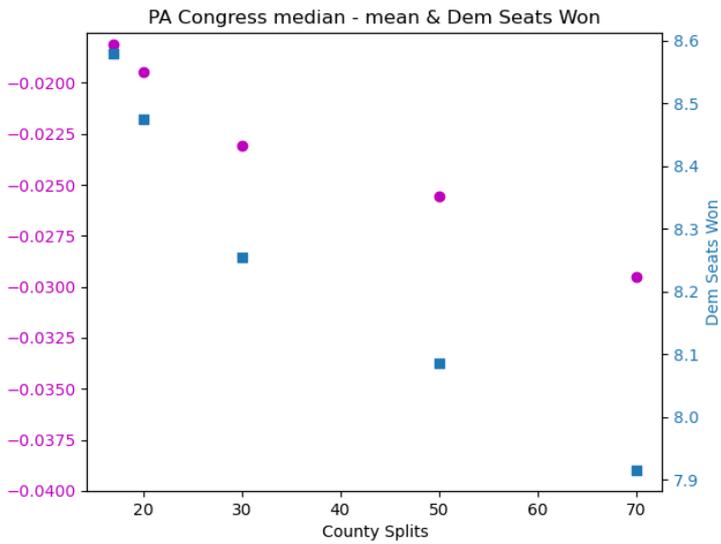
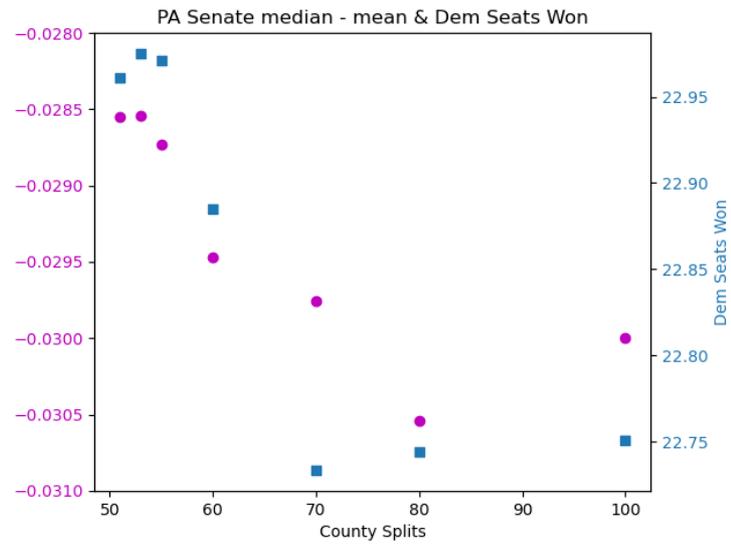

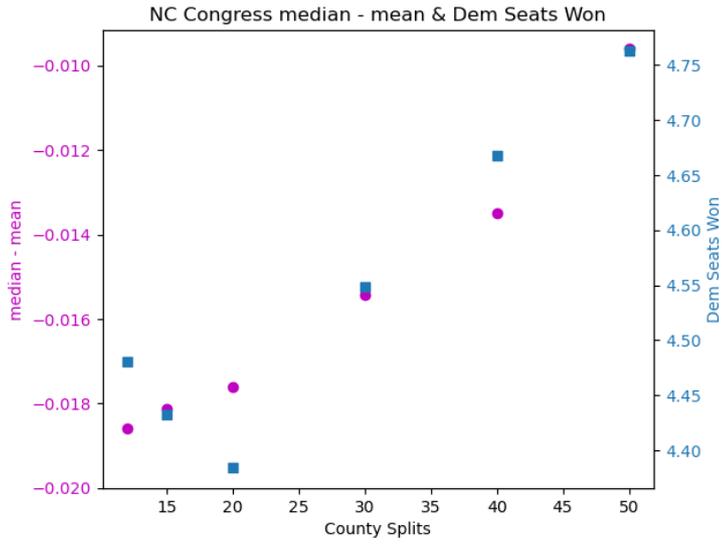
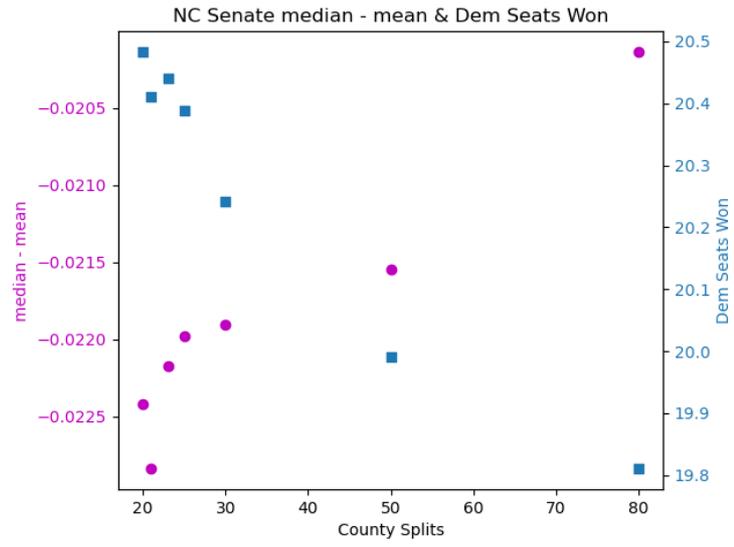
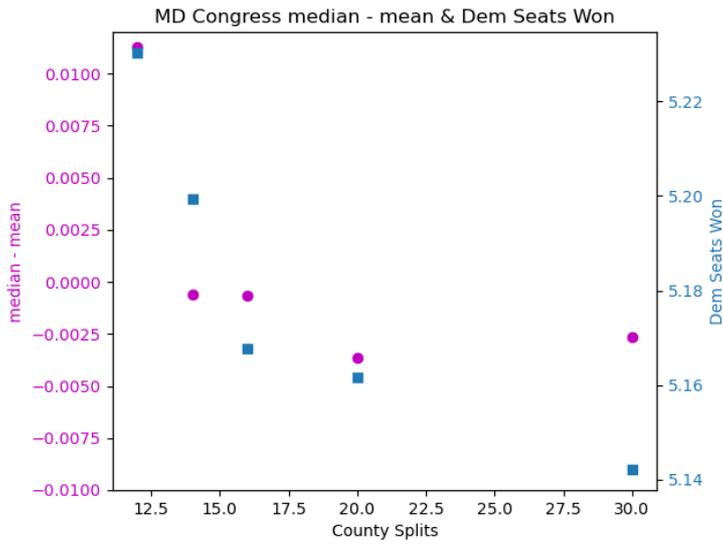
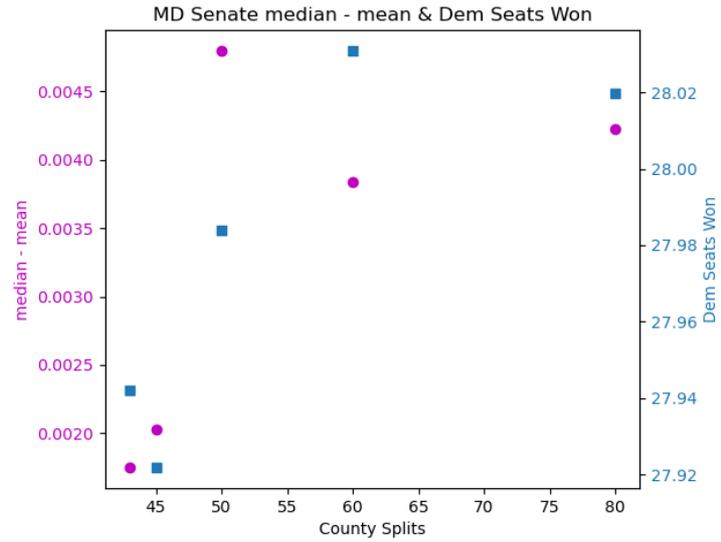

**Conclusions**

We show that VTD (precinct) histograms of partisanship by state are a powerful but simple tool for predicting mean values of seats won in Markov chain simulations for State Senate and Congressional districting, covering a 10-state study. The fraction of equal-voter weighted VTD's won is a good proxy for the fraction of seats won, at least for simulations. Histograms of partisanship reveal large skew with Democrats prevailing by large margins in urban districts, not outweighed by Republican wins in rural areas. The values for skew ranged from 0.16 for Maryland to 0.98 in Michigan. These values for skew are related to large values for [median – mean].

High degrees of skew, apart from any gerrymandering, account for a significant Republican advantage as even when votes are evenly split statewide, Democratic concentration in cities distributes votes inefficiently. Sometimes, instead of receiving a "loser's bonus" by getting at least some seats while achieving < 50% of the statewide vote, it is possible to achieve more than 50% of the seats with < 50% of the statewide vote. The VTD share analysis shows this effect, with smaller departures from mean simulation results for State Senate seats than for Congressional seats, where the variances in population distribution are relatively lower.

Variations in simulation results, apart from mean values are considerable, with efficiency gap ranges up to 20% for State Senate and 30% for Congressional seats. While mean values for metrics and simulated seats won are sometimes held to be the unbiased value for comparison with a gerrymandered district plan, often gerrymanders are outside the pale of any distributions. With such wide ranges within random simulations however, we raise the question whether an unbiased districting plan is one where the median is close to the mean of the simulation ensemble. That is, selecting a simulation result where the median VTD is close to the mean VTD may achieve a more equitable representation apart from scenarios closest to mean values over simulation scenarios where partisanship exhibits a skewed geographical distribution.

County splits are a key metric for preventing extravagant gerrymanders, with districts snaking through to all four corners of a State, as well as a tool complementing compactness requirements. When constraining simulations to smaller numbers of county splits to probe effect on bias, we find mixed results with some states showing lower [median – mean] metrics with fewer county splits but others showing greater. Pennsylvania appears to show some reduction in bias with fewer county splits but limits on county splits do not substitute for rules that reduce partisan bias more directly. While distorted maps are an easy 'tell' for gerrymanders easy to expose in public, sophisticated gerrymanders present compact maps (eg. the Turzai-Scarnati Pennsylvania plan). Maps with the most equitable representation, translating votes into seats the most efficiently, may not be the most compact or represent the mean outcomes from large simulation ensembles.

**Acknowledgements**

We thank Darryl DeFord for his invaluable help on the gerrychain software, and Alec Ramsay of Daves' Redistricting for comments as well as Florida election data. We thank John Nagle for his comments on manuscript.

**Declaration of Interests**

We have no financial interests to declare.

**References**

Bishop, W. (2009), "The Big Sort," Boston: Houghton Mifflin

Chen, J., Rodden, J. (2013), "Unintentional Gerrymanding: Political Geography and Electoral Bias in Legislatures," Quarterly Journal of Political Science 8: 239-269

Deford, D., Duchin, M., Solomon, J. (2019), "ReCombination: A Family of Markov Chains for Redistricting," arXiv no. 1911.05725. [70,71,76,77]

Duchin, M. (2018), in Politifact, https://www.politifact.com/factchecks/2018/jan/10/robin-hayes/nc-gop-gerrymander-strange-looking-monster-drawing/

Duchin, M., Gladkova, T., Henninger-Voss, E., Klingensmith, B., Newman, H., Wheelen, H. (2019) "Locating the Representational Baseline: Republicans in Massachusetts," Election Law Journal 18: 388-401

Kendall, M. G. and Stuart, A, (1950), "The Law of the Cubic Proportion in Election Results," British Journal of Sociology 1(3): 183-186

Nagle, J., (2019), "What Criteria Should Be Used For Redistricting Reform," Election Law Journal 18: 63-77

Rodden, J. A. (2019) ,"Why Cities Lose," New York: Basic Books